\documentclass[sigconf]{acmart}

\begin{document}

\copyrightyear{2026}
\acmYear{2026}
\setcopyright{cc}
\setcctype{by}
\acmConference[ICMR '26]{International Conference on Multimedia Retrieval}{June 16--19, 2026}{Amsterdam, Netherlands}
\acmBooktitle{International Conference on Multimedia Retrieval (ICMR '26), June 16--19, 2026, Amsterdam, Netherlands}
\acmDOI{10.1145/3805622.3810623}
\acmISBN{979-8-4007-2617-0/2026/06}

\title{Event-Based Token Sequences for Audio-Conditioned Music-Game Level Modeling}

\author{Ke Zhang}
\authornote{Corresponding author.}
\orcid{0009-0009-4937-1252}
\affiliation{%
  \institution{Japan Advanced Institute of Science and Technology}
  \city{Nomi}
  \state{Ishikawa}
  \country{Japan}}
\email{s2660002@jaist.ac.jp}

\author{Chu-Hsuan Hsueh}
\orcid{0000-0001-8888-3116}
\affiliation{%
  \institution{Japan Advanced Institute of Science and Technology}
  \city{Nomi}
  \state{Ishikawa}
  \country{Japan}}
\email{hsuehch@jaist.ac.jp}

\author{Kokolo Ikeda}
\orcid{0009-0007-0566-8172}
\affiliation{%
  \institution{Japan Advanced Institute of Science and Technology}
  \city{Nomi}
  \state{Ishikawa}
  \country{Japan}}
\email{kokolo@jaist.ac.jp}

\begin{abstract}
Procedural generation of music game levels is an exciting yet challenging problem, as levels must translate musical structure into interactive sequences of timed gameplay events.
Most existing approaches formulate this task by frame-based representations, dividing audio into uniform time grids and predicting events at each frame.
This makes gameplay events implicit across many frames. As a result, it is hard to describe event-level timing relations and longer-range structure found in human-authored levels.
We use procedural generation as a practical setting to study how musical cues map to interactive event sequences.
Inspired by event-based symbolic music modeling, we propose a token-level sequence formulation that casts level generation as a multimodal sequence-to-sequence problem.
Conditioned on an audio excerpt and level metadata, the model generates a token sequence alternating gameplay-event and beat-shift tokens. This explicitly represents actions and their relative timing in beat space.
Based on this formulation, we build a Transformer model. It outperforms representative frame-level baselines under event-level evaluation. It also enables systematic analysis of how audio supports rhythm-aligned event prediction beyond metadata conditioning.
\end{abstract}

\ccsdesc[500]{Applied computing~Sound and music computing}
\ccsdesc[300]{Information systems~Music retrieval}
\ccsdesc[300]{Computing methodologies~Neural networks}

\keywords{music game, rhythm game, multimodal sequence modeling, token-level modeling, audio-conditioned generation}

\maketitle

\section{Introduction}
Understanding how musical cues map to structured event sequences is a core challenge in music-game level modeling, and relates to broader questions in music information retrieval on representing and interpreting audio~\cite{schedl2014mir}.
Procedural content generation (PCG) provides a practical setting to test this audio--event correspondence, because the mapping must be realized as playable interactive content~\cite{summerville2018pcgml}.
A music game level is not merely a sequence of actions, but a temporally structured artefact shaped by musical rhythm, phrasing, and difficulty design.
Well-designed levels translate musical structure into expressive and playable event sequences, requiring both precise local timing and coherent long-range organization.
Generating such content demands models that bridge continuous audio signals and discrete symbolic actions, while preserving temporal structure across multiple scales.
At its core, level modeling requires identifying salient temporal cues from rich, time-varying audio and organizing them into structured event sequences that players can directly interact with.
This perspective emphasizes event patterns, rather than dense uniform time grids.
While the objective is to synthesize new interactive structures, doing so in a musically grounded and playable manner depends critically on how audio--event relationships are represented and modeled.
As a result, the choice of temporal representation and sequence modeling formulation plays a central role in determining generation quality.

Most existing approaches to automatic music game level modeling adopt frame-based formulations~\cite{donahue2017dance,liang2019procedural,takada2023genelive}.
These methods discretize time into uniform grids and perform frame-level classification or onset detection on audio signals, sometimes augmented with beat-synchronous guidance~\cite{takada2023genelive,goto2001realtime,ellis2007beat}.
While effective for local event detection, frame-based representations treat gameplay events as implicit patterns distributed across frames.
As a consequence, event-to-event timing relationships, beat-relative offsets, and long-range temporal regularities are not modeled explicitly, but must be inferred indirectly from sequences of dense frame-level predictions.
This often makes it difficult to capture the structural properties that are central to human-authored game levels, such as rhythmic consistency, phrasing, and controlled variation across difficulty levels.
In particular, these formulations do not provide an explicit representation of event-level temporal structure, which is central to rhythmic consistency and level design.

Similar challenges have been observed in other temporally structured domains.
In symbolic music modeling, early piano-roll or frame-based representations were found to be limited in their ability to capture expressive timing and hierarchical musical structure~\cite{boulanger2012modeling}.
Motivated by this, recent work has increasingly shifted toward event-based symbolic representations, in which musical content is modeled as sequences of discrete events with explicit relative timing~\cite{oore2018time,huang2018music,huang2020pop}.
Event-based sequence modeling, particularly when combined with transformer architectures, has been shown to better capture long-range dependencies, rhythmic regularity, and structural coherence in music generation~\cite{huang2018music,huang2020pop}.
These advances provide an important conceptual foundation for rethinking how temporal structure should be represented in music game level modeling.

Inspired by event-based symbolic music modeling, we formulate music game level generation as a token-level sequence modeling problem.
In this work, we introduce a token-level music game level reconstruction task that casts level modeling as a multimodal sequence-to-sequence problem.
Given a short audio excerpt and level-related metadata such as beats per minute (BPM) and difficulty, the model predicts a sequence of discrete tokens consisting of beat-shift tokens and gameplay event tokens.
This representation operates directly at the level of events and relative timing, allowing transformer-based models to reason over gameplay structure in a manner analogous to symbolic music generation~\cite{chan2016listen,vaswani2017attention}, while remaining grounded in the interactive constraints of music games.

An important aspect of this formulation is how audio complements metadata and sequence priors. 
While metadata and autoregressive priors provide strong structural cues, accurate event placement and articulation ultimately depend on musical timing information conveyed by audio.
Disentangling the contribution of audio from that of metadata and sequence modeling remains non-trivial in multimodal temporal generation tasks~\cite{goto2017miroverview,casey2008content}.
To systematically analyze this interaction, we conduct controlled ablation and perturbation experiments and introduce a dedicated evaluation metric, the Audio Contribution Score (ACS), which quantifies the relative contribution of audio beyond metadata conditioning and sequence priors under a fixed evaluation protocol.

Our contributions are summarized as follows:
\begin{itemize}
    \item We formulate music game level modeling as token-level sequence modeling conditioned on audio and metadata, emphasizing event-based representations for capturing rhythmic structure, inter-event timing, and longer-range dependencies in interactive content.
    \item We propose a multimodal sequence-to-sequence framework that predicts beat-shift and gameplay event token sequences from audio and level metadata, enabling transformer-based models to operate directly at the event level.
    \item We conduct systematic ablation and perturbation experiments to analyze the role of audio in level reconstruction, and introduce the Audio Contribution Score (ACS) to quantify the independent contribution of audio to generation performance under controlled settings.
\end{itemize}

\section{Related Work}

Early work on music-game level modeling focused on mapping musical cues to gameplay actions through onset- or beat-driven heuristics. Foundational beat-tracking research, such as Goto’s real-time estimator~\cite{goto2001realtime} and Ellis’s dynamic-programming formulation~\cite{ellis2007beat}, provided mechanisms for aligning musical structure with gameplay timing. With the adoption of deep learning, frame-level prediction approaches emerged: Dance Dance Convolution (DDC)~\cite{donahue2017dance} employed CNN–RNN architectures to classify note actions over dense temporal discretisations, while later work explored richer acoustic features, beat-synchronous guidance, and difficulty control within similar frame-based formulations~\cite{liang2019procedural,lin2019generationmania,takada2023genelive,tsujino2018gradation}. These approaches operate over fixed temporal grids and emphasize stable alignment between musical structure and gameplay timing.

Event-based representations are widely used in symbolic music modelling. Early RNN-based models demonstrated that event sequences capture expressive timing and polyphonic dependencies more naturally than piano-roll grids~\cite{boulanger2012modeling,hochreiter1997lstm}. PerformanceRNN~\cite{oore2018time} introduced time-shift tokens to explicitly model relative timing, while Transformer-based models such as Music Transformer~\cite{huang2018music} and MuseNet~\cite{payne2019musenet} further showed the effectiveness of event-based tokenisation for long-range structural modelling. Subsequent work proposed structured vocabularies that integrate metrical context, such as REMI~\cite{huang2020pop}, and explored symbolic abstraction through GAN-based models, hierarchical latent-variable formulations, and convolutional approaches~\cite{dong2018musegan,roberts2018hierarchical,huang2016counterpoint,hadjeres2017deepbach}.

Multimodal sequence modeling has commonly been studied using encoder–decoder architectures with pretrained audio encoders. Whisper~\cite{radford2022whisper} provides robust timing-sensitive representations learned from large-scale weakly supervised data, while MERT~\cite{li2024mert} offers music-oriented embeddings capturing spectral, harmonic, and structural characteristics. Classical attention-based encoder–decoder models such as Listen, Attend and Spell~\cite{chan2016listen,chan2015las}, together with alignment-free sequence objectives like CTC~\cite{graves2006ctc}, established audio-to-token modeling paradigms. Transformer-based architectures originating from language representation learning~\cite{vaswani2017attention,devlin2019bert} have since become a standard backbone for multimodal generation, and recent audio language models such as AudioLM~\cite{borsos2022audiolm} further demonstrate the effectiveness of discrete token modeling across modalities~\cite{srivastava2012multimodal,bengio2013representation}.

\section{Problem Formulation}

\subsection{Task Definition}

At the task level, the task of music-game level modeling is to generate a complete gameplay level from a music track and associated level metadata, such as tempo and difficulty.
This task definition is independent of specific modeling or implementation choices, and characterizes the underlying problem shared by different level generation approaches.

\subsection{Modeling Assumptions and Formulation}
\label{sec:modeling_assumptions}

In this work, we adopt a multimodal sequence modeling formulation to address this task.
To make training and inference computationally tractable, we operate on audio excerpts \(a\) and the corresponding level segments; the formulation itself does not depend on a specific segmentation strategy. Segmentation is performed in beat space, with each excerpt covering a globally consistent number of beats across the dataset.
Each audio excerpt is processed by an audio feature extractor (e.g., a pretrained audio encoder) to obtain a sequence of acoustic representations
\[
x = f_{\text{audio}}(a),
\]

We assume that level-related metadata
\[
m = \{\text{BPM}, \text{skill rating}, \text{difficulty tier}\}
\]
is given as input, following common human level design practice.
In particular, BPM defines a global beat-space reference used to express gameplay timing.
Here, skill rating represents a continuous measure of absolute technical difficulty assigned by human designers, while difficulty tier denotes a discrete level category (e.g., beginner, expert) that distinguishes multiple level variants of the same music track.
Skill rating reflects overall technical load. Difficulty tier reflects chart writing style, which tends to become more elaborate at higher tiers. The two are treated as separate conditioning signals.

At a high level, the goal is to model the conditional distribution
\[
p(y \mid x, m),
\]
where $x$ denotes the acoustic representation extracted from an audio excerpt, and $y$ denotes the event sequence of the corresponding level segment.

In this work, we realize this conditional distribution using an
autoregressive sequence modeling formulation.
Specifically, the joint distribution is factorized as
\[
p(y \mid x, m)
= \prod_{i=1}^{N} p(y_i \mid y_{<i}, x, m),
\]
where each token \( y_i \) represents a discrete event in the target sequence.

\subsection{New Event-Based Representation of Music Game Level}
\label{sec:beat_shift_representation}

\begin{figure}[t]
  \centering
  \includegraphics[width=0.5\textwidth]{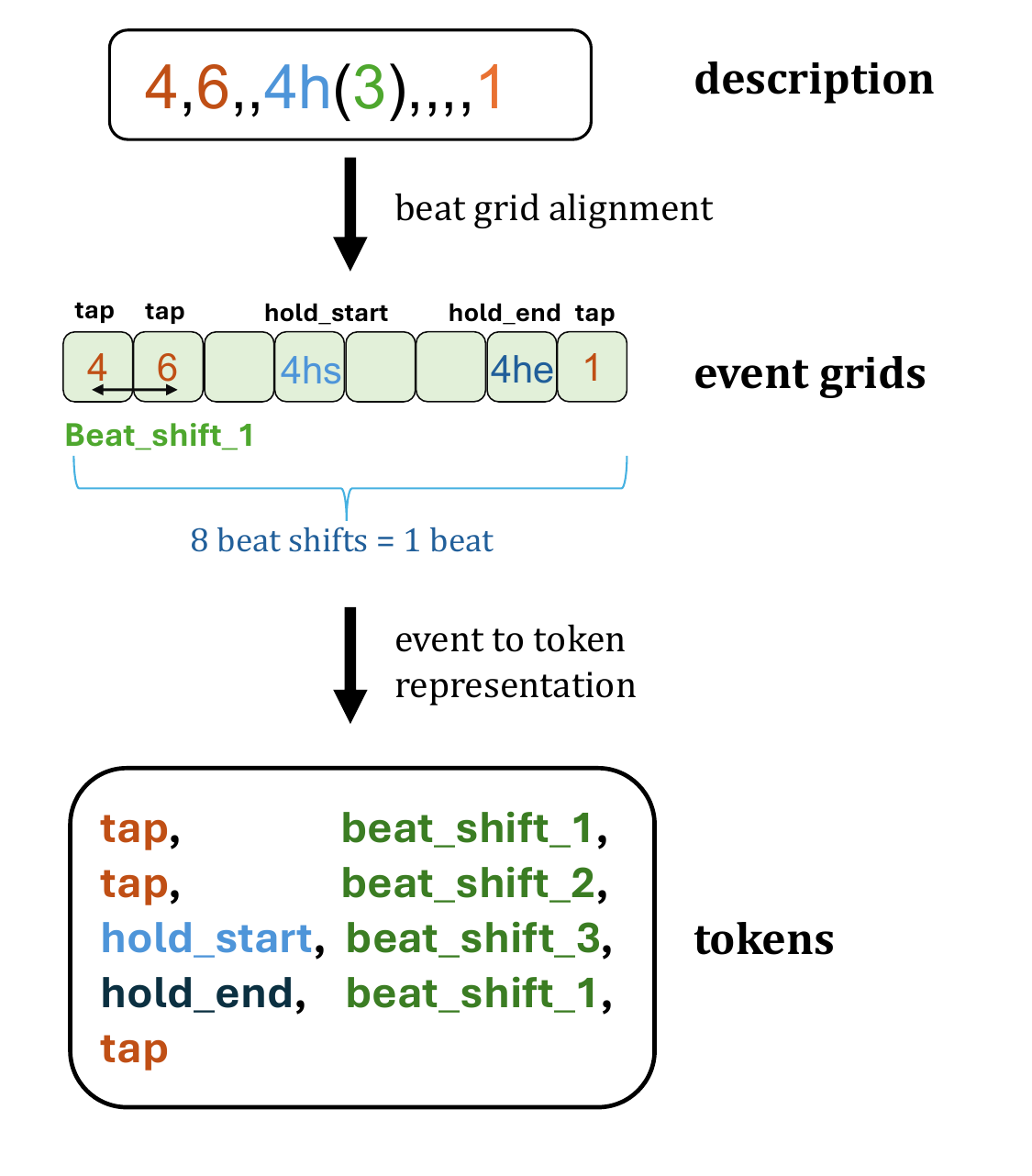}
  \caption{Illustration of the proposed event-based representation construction.
A beat-aligned raw level description is interpreted as an ordered sequence of gameplay events in beat space.
Temporal progression between consecutive events is encoded using beat-shift tokens, while event semantics are encoded using event tokens from a predefined vocabulary.
The resulting token sequence alternates between beat-shift and event tokens, forming a linear representation suitable for autoregressive sequence modeling.
The example visualizes how discrete gameplay events and musically grounded timing information are jointly captured by the proposed representation.}
  \Description{Three-panel vertical diagram illustrating the event-based representation. The top panel shows a raw simai-style level description string ``4,6,,4h(3),,,,1''. A downward arrow labeled ``beat grid alignment'' leads to the middle panel, an 8-cell event grid spanning one beat (annotated ``8 beat shifts = 1 beat''), with cells filled by tap events at positions 4 and 6, a hold\_start at position 4 (``4hs''), a hold\_end at position 4 (``4he''), and a trailing tap at position 1. A second downward arrow labeled ``event to token representation'' leads to the bottom panel, a token sequence box listing alternating event and beat-shift tokens: tap, beat\_shift\_1, tap, beat\_shift\_2, hold\_start, beat\_shift\_3, hold\_end, beat\_shift\_1, tap.}
  \label{fig:event_representation}
\end{figure}

Designing an effective representation for music-game levels requires capturing the symbolic structure of gameplay events.
It also requires capturing the musically grounded timing that governs their placement.

We follow the community-adopted simai maidata format, which encodes events using beat-relative offsets~\cite{simai_maidata_spec}.
Motivated by this representation, we abstract each level segment as a linear sequence of discrete tokens for modeling purposes.
The sequence alternates between beat-shift tokens and event tokens; Figure~\ref{fig:event_representation} provides an overview of the representation construction process.
This formulation enables event-centric sequence modeling with explicit temporal semantics.
It is naturally compatible with Transformer-based generative models.

\paragraph{Event Vocabulary}
The event-based representation is defined at an abstract level.
In this work, the specific event vocabulary is instantiated for touch-based rhythm games such as \textit{maimai}.
Aside from special tokens (<pad> for padding, <s> and </s> for sequence start and end, and <unk> for unknown tokens), the vocabulary contains five classes of gameplay events: tap, hold (start/end pairs), slide, double (simultaneous taps), and break (accented or high-impact actions).
This coarse yet expressive set covers the gameplay semantics sufficient for the target reconstruction setting, while avoiding an explosion in fine-grained gesture variants.

Hold actions are represented using paired start and end events, which captures duration while maintaining a simple event-centric sequence structure.
Overlapping holds are represented as start/end events without explicit object IDs.
In practice, we recover start--end pairing deterministically during reconstruction by matching each hold-end to the most recent unmatched hold-start.
In our dataset, this rule was sufficient for reconstruction and we did not observe unmatched hold-start events.
The model predicts only these symbolic event tokens; timing and event types are deterministically reconstructed from the token sequence under the above decoding rules.

Raw level descriptions in many rhythm games include additional positional attributes, such as lane indices or spatial coordinates.
In prior work on music-game level modeling, including systems targeting osu! or DDR-style levels, event generation is typically formulated as a two-stage process: predicting whether a gameplay event occurs at a given temporal location and then assigning positional or lane-specific attributes.
In this work, we revise the first-stage formulation to account for datasets with multiple distinct action types, modeling event prediction as a multi-class problem.
For consistency with existing evaluation protocols and fair comparison at the event-sequence level, we focus on this revised first-stage formulation and model only the occurrence and symbolic type of gameplay events.
Accordingly, positional information is intentionally excluded from the proposed token representation.
This design choice reflects the scope of the current study rather than a limitation of the representation framework itself; extensions to incorporate positional attributes are left for future work.

\paragraph{Beat-Shift Timeline (Beat-Space Timing)}
Temporal progression is represented using beat-based offsets, consistent with rhythm-game authoring in beat space.
Unlike MIDI-style symbolic modeling, which often measures time in milliseconds or ticks, music-game levels are authored natively in musical beats.
Gameplay timing, including rhythmic density, off-beat accents, and difficulty scaling, is therefore defined relative to tempo rather than absolute time.
Rather than adopting a dense beat-aligned grid that requires many empty symbols to represent temporal gaps, beat-shift tokens encode event-to-event timing directly.

Let the current beat position be $b$.
Each beat-shift token $k$ advances the cursor by $\Delta b = k / U$, where $U$ denotes the number of units per beat ($U = 8$ in our implementation).
This formulation operates in beat space, consistent with simai maidata chart specifications, and avoids lossy conversion to frame indices.
The resulting token sequence alternates between beat-shift and event tokens, enabling deterministic reconstruction of level timing by accumulating beat offsets and placing events at the resulting beat positions.

\paragraph{Representation Construction}
The input level description follows simai maidata chart specifications and is represented as a beat-aligned discrete temporal structure, with optional gameplay events at each time unit.
Under this assumption, the construction of the event-based representation proceeds in three steps.
First, the beat-aligned description is interpreted as an ordered sequence of gameplay events, together with their relative positions in beat space.
Second, temporal progression between consecutive events is encoded using beat-shift tokens, while the event semantics themselves are encoded using event tokens from the predefined vocabulary.
Finally, the resulting token sequence alternates between beat-shift and event tokens, forming a linear representation suitable for autoregressive modeling.

Figure~\ref{fig:event_representation}
illustrates this process with a concrete example.
In the figure, a beat-aligned raw level description is first mapped to a symbolic event sequence defined in beat space, and then deterministically converted into an event-based token sequence by inserting beat-shift tokens to represent relative timing between successive events.
This example serves to visualize how discrete gameplay events and musically grounded timing information are jointly captured by the proposed representation, rather than to define any specific file format or preprocessing rule.

\paragraph{Advantages Over Frame-Based Representations}
Compared to frame-based approaches, including beat-aware onset detection methods where beat information is used as auxiliary guidance, the proposed representation offers the following advantages:
(i) it adopts an event-centric formulation that models gameplay as a sequence of discrete events rather than dense frame-wise predictions;
(ii) it represents temporal progression explicitly through event-to-event relations, instead of relying on an external alignment signal;
(iii) by operating directly in beat space, it preserves musically grounded timing semantics aligned with rhythm-game design practice;
(iv) it expresses event-to-event timing directly in beat space, avoiding long runs of empty symbols on a dense grid; and
(v) it supports a direct, rule-based mapping from tokens to events, without relying on additional smoothing or peak-detection heuristics.

\section{Proposed Framework}

\begin{figure*}[!t]
  \centering
  \includegraphics[width=2.0\columnwidth]{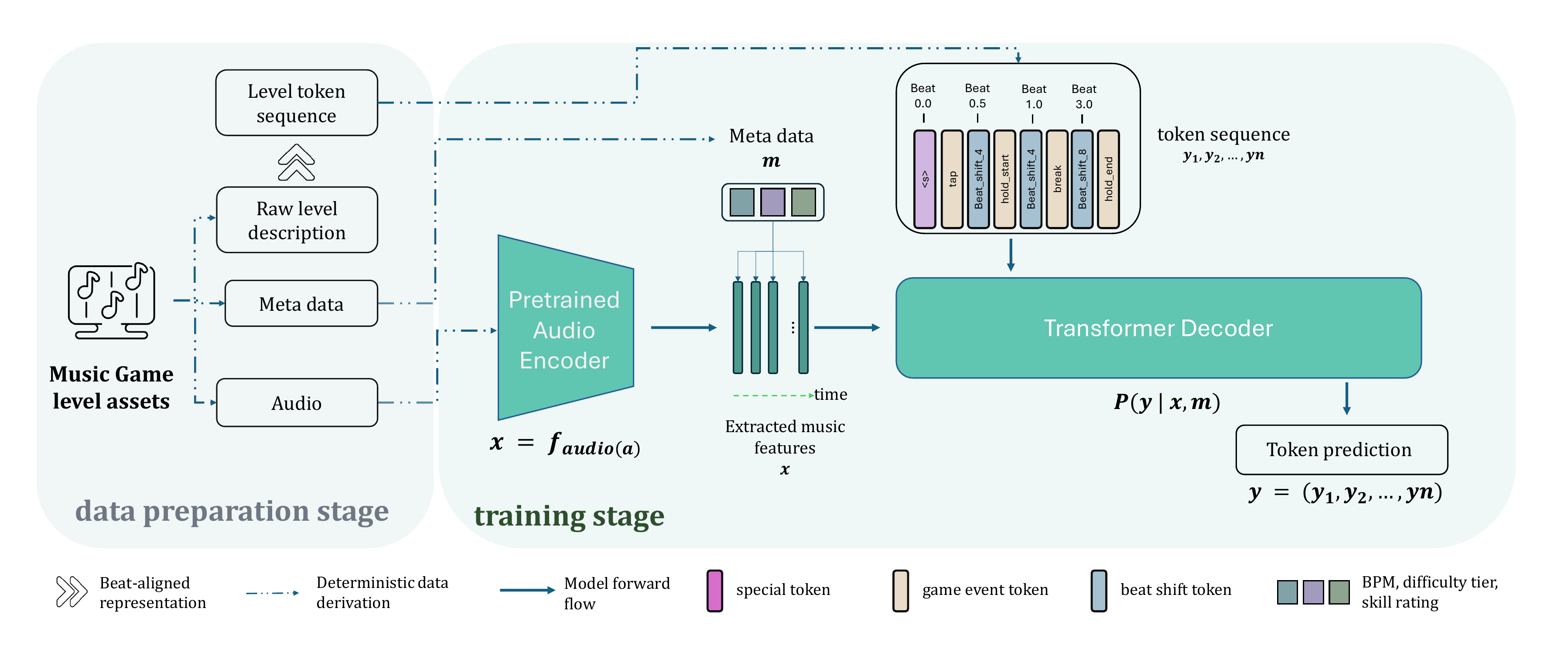}
  \caption{Overview of the proposed pipeline.
  Left: deterministic data preparation from commercial music-game assets, deriving beat-aligned level representation, metadata \(m\) (e.g., BPM, difficulty tier, skill rating), and the supervision token sequence \(y\).
  Right: training-stage forward flow, where a pretrained audio encoder extracts acoustic features \(x=f_{\text{audio}}(a)\), which are conditioned on \(m\) and fed to an autoregressive Transformer decoder to model \(p(y_t \mid y_{<t}, x, m)\) using teacher forcing.
  Tokens include special tokens, gameplay-event tokens, and beat-shift tokens.}
  \Description{Wide two-panel horizontal flow diagram. The left panel, labeled ``data preparation stage'', shows a music-game level assets icon branching into three parallel boxes---raw level description, metadata, and audio---connected by dash-dotted arrows denoting deterministic derivation; a double-chevron arrow converts the raw level description into a beat-aligned representation and then into the supervision ``level token sequence'' box. The right panel, labeled ``training stage'', shows the audio stream entering a trapezoidal ``Pretrained Audio Encoder'' that outputs a sequence of frame-level feature vectors \(x = f_{\text{audio}}(a)\) along a time axis. A small metadata vector \(m\) (three colored squares representing BPM, difficulty tier, and skill rating) merges with these features and feeds into a large rectangular ``Transformer Decoder'' block. Above the decoder, a horizontal token sequence strip shows example tokens \(\langle s\rangle\), tap, beat\_shift\_4, hold\_start, beat\_shift\_4, break, beat\_shift\_8, hold\_end aligned to beat positions 0.0, 0.5, 1.0, and 3.0. The decoder's downward output arrow is labeled \(P(y \mid x, m)\), leading to a ``token prediction'' box with \(y = (y_1, y_2, \ldots, y_n)\). A legend at the bottom distinguishes beat-aligned representation chevrons, deterministic data derivation arrows, model forward-flow arrows, and color-coded token categories (special, game-event, beat-shift).}
  \label{fig:pipeline_overview}
\end{figure*}

Our model adopts an encoder--decoder architecture that maps an audio excerpt and associated level metadata to a sequence of beat-shift and gameplay-event tokens.
The overall formulation, from deterministic data preparation to training-stage sequence modeling, is illustrated in Figure~\ref{fig:pipeline_overview}.
The framework consists of a pretrained audio encoder, a metadata embedding module, and an autoregressive Transformer decoder.

\subsection{Data Preparation from Game Assets}

The data preparation stage, illustrated in the left part of Figure~\ref{fig:pipeline_overview}, defines a deterministic pipeline for constructing all training inputs and supervision targets from raw music-game assets.

In commercial rhythm games, a gameplay level is specified by a level package that defines how a player interacts with a given music track.
In this work, we assume that each package includes a raw level description aligned in beat space, level-specific metadata (e.g., BPM and difficulty), and a reference to an associated audio track.

The raw level description is first parsed to obtain an ordered sequence of gameplay events.
Following the event-based formulation introduced in Section~\ref{sec:beat_shift_representation}, this event sequence is deterministically converted into an event-based token sequence consisting of beat-shift tokens and gameplay-event tokens.
This token sequence serves as the supervision target for sequence modeling.

Level metadata, including BPM, skill rating, and difficulty tier, are extracted directly from the level specification and used as global conditioning signals.
The associated audio track is retrieved as an external resource and processed by the audio encoder described in Section~\ref{sec:audio_encoder}.
All experiments in this work use training samples constructed through the same deterministic data preparation pipeline.

\subsection{Audio Encoder}
\label{sec:audio_encoder}

Rather than relying solely on fixed-resolution mel-spectrogram features, we use a pretrained audio encoder—either Whisper or MERT—to obtain temporally contextualized acoustic representations from the input audio excerpt.
Both encoders produce frame-level acoustic representations at a temporal resolution of approximately 50~Hz (corresponding to a hop size of 20--25~ms), yielding a sequence
\[
h = (h_1, \ldots, h_T), \quad h_t \in \mathbb{R}^{d_{\text{model}}},
\]
which is used as the key/value memory for cross-attention in the decoder.
We do not enforce explicit beat–frame alignment; the decoder attends to frame-level acoustic representations via cross-attention.

\subsection{Metadata Embedding and Fusion}
\label{sec:metadata_embedding}

Metadata consists of BPM, skill rating, and difficulty tier, which provide global structural cues for level generation.
In the context of music-game level authoring, and following common practice in community-created content (e.g., simai), we assume that a global tempo (BPM) is available and focus on generating gameplay events relative to this temporal reference.

While a song may contain local tempo variations or multiple tempo sections, level files typically specify a single canonical BPM, often chosen as a common divisor that allows all gameplay events to be expressed using integer or simple fractional beat offsets.
This BPM therefore serves as a unified temporal scale for describing event spacing across the entire level.
As defined in Section~\ref{sec:modeling_assumptions}, skill rating and difficulty tier represent complementary notions of difficulty.

\paragraph{Scalar metadata (BPM, skill rating)}
BPM and skill rating are treated as scalar values and projected into the model space through independent learnable linear transformations:
\[
b_{\text{emb}} = \text{Linear}_{\text{bpm}}(\text{BPM}), \quad
d_{\text{emb}} = \text{Linear}_{\text{diff}}(\text{rating}).
\]

\paragraph{Categorical metadata (difficulty tier)}
The difficulty tier is treated as a categorical variable with five discrete levels and embedded using a dedicated learnable embedding:
\[
t_{\text{emb}} = \mathrm{Embedding}_{5 \times d_{\text{tier}}}(\mathrm{tier}).
\]

\paragraph{Fusion}
We inject metadata by broadcasting a single global metadata vector to all encoder time steps and concatenating it with the frame-level acoustic representations.
Specifically, we obtain a metadata vector by concatenating the three embeddings,
\[
m_{\text{enc}} = [\, b_{\text{emb}} \,;\, d_{\text{emb}} \,;\, t_{\text{emb}} \,],
\]
and apply it to every encoder frame by
\[
h'_t = [\, h_t \,;\, m_{\text{enc}} \,], \quad t=1,\ldots,T,
\]
where $[\,\cdot\,;\,\cdot\,]$ denotes concatenation and $m_{\text{enc}}$ is shared across all time steps.
This fusion strategy appends global information about tempo, difficulty, and level style to the acoustic representations used for decoding, consistent with the conditioning pathway illustrated in Figure~\ref{fig:pipeline_overview}.
We do not prepend metadata tokens to the decoder input.

\subsection{Autoregressive Decoder}
\label{sec:autoregressive_decoder}

The decoder generates a temporally ordered sequence of gameplay events conditioned on both the acoustic representations produced by the encoder and global level metadata.
This corresponds to the training-stage forward flow shown in the right part of Figure~\ref{fig:pipeline_overview}.
To model long-range dependencies over heterogeneous token types while allowing flexible alignment between audio cues and symbolic events, we adopt an autoregressive sequence modeling formulation.

We use a standard Transformer decoder with learned positional embeddings, causal self-attention, and multi-head cross-attention to the encoder's frame-level acoustic representations; exact hyperparameters are reported in Section~\ref{sec:implementation}.

The output vocabulary consists of beat-shift tokens, representing relative temporal offsets in beat space (as defined in Section~\ref{sec:beat_shift_representation}), and gameplay-event tokens, including tap, hold, slide, double, and break.
Although beat-shift and gameplay-event tokens alternate in principle, this alternation is not enforced during training, in order to avoid imposing a hard structural constraint on the sequence model.

\subsection{Inference}

Autoregressive decoding uses either greedy decoding or beam search with beam size 5 (default); length normalization during beam search mitigates locally optimal but globally inconsistent outputs.

Decoding terminates when the end-of-sequence token \texttt{</s>} is produced.
In addition, we enforce a hard termination constraint based on temporal coverage.
During decoding, beat-shift tokens are cumulatively summed to track the generated temporal length in beat space.
If the accumulated beat-shift exceeds the total beat length of the input audio excerpt, decoding is forcibly terminated.

The total beat length is defined by the fixed-length beat-space segmentation of the input audio, as described in Section~\ref{sec:modeling_assumptions}.
Decoded token sequences are deterministically converted into event-level symbolic descriptions by accumulating beat-shift values and placing gameplay events at the resulting beat positions.

\section{Evaluation}

\subsection{Datasets}
\label{sec:datasets}

The level data used in this study were extracted from a commercially released rhythm game (maimai) and were used solely for academic research purposes.
The corresponding audio tracks and original level files are not redistributed.

Due to copyright restrictions, we do not release any data that would allow reconstruction of the original game levels.
Instead, we release illustrative, non-invertible token sequence examples on a public repository, together with a formal specification of the event vocabulary and beat-shift rules.
These example sequences are provided solely to demonstrate the proposed representation format and do not correspond to any real in-game levels.
In addition, we release the full data conversion and training code used in our pipeline, enabling other researchers to apply the proposed methodology to their own legally obtained data.

Separately from data availability considerations, we define the following experimental protocol.
The level data consist of 4,187 levels from 1,018 unique songs, obtained through standard gameplay and legally accessible channels.
Levels are split into training, validation, and test sets using a 70/15/15 ratio at the song level, ensuring that no song appears in more than one split.
This results in 3,350 training levels, 421 validation levels, and 416 test levels.

Each chart is further segmented into fixed-length excerpts in beat space, yielding excerpt-level token sequences used for model training and evaluation.
The resulting numbers of sequences are 21,595 for training, 2,808 for validation, and 2,669 for testing.
All statistics reported below are computed at the excerpt (sequence) level unless otherwise noted.
Table~\ref{tab:dataset_stats} summarizes key characteristics of the resulting tokenized sequences.
The token vocabulary consists of five gameplay event types and beat-shift tokens representing relative temporal offsets in beat space (Section~\ref{sec:beat_shift_representation}).

\begin{table}[htbp]
\centering
\caption{Statistics of the excerpt-level token sequences used in the experiments.}
\label{tab:dataset_stats}
\begin{tabular}{l p{0.3\columnwidth}}
\toprule
Statistic & Value \\
\midrule
Number of songs & 1,018 \\
Number of levels & 4,187 \\
Number of sequences (train / valid / test) & 21,595 / 2,808 / 2,669 \\
Avg. tokens per sequence & 120--260 \\
Beat-shift tokens & $\sim$54\% \\
Event tokens & $\sim$46\% \\
Max sequence length & 1,024 tokens \\
\bottomrule
\end{tabular}
\end{table}

Beat-shift tokens dominate in lower-difficulty excerpts, while temporal event density increases consistently with chart difficulty.

\subsection{Implementation Details}
\label{sec:implementation}

We use either Whisper-base or MERT as the audio encoder backbone in separate experiments.
Unless otherwise stated, all model components, including the audio encoder, Transformer decoder, and metadata embedding modules, are optimized jointly from the first training step without freezing.
In all experiments, each audio excerpt spans 80 beats, corresponding to the fixed beat-length segmentation described in Section~\ref{sec:modeling_assumptions}.
No additional pooling, convolutional prenet, or temporal downsampling is applied, and the temporal resolution is determined solely by the underlying pretrained encoder.
The decoder is a 12-layer Transformer with a hidden dimension of 768, a feed-forward dimension of 2048, and 8 attention heads per layer.
The token vocabulary and beat-shift representation follow the definitions introduced in Section~\ref{sec:beat_shift_representation}.

Training uses standard teacher-forced cross-entropy loss over the target token sequence, with label smoothing set to 0.05.
All decoder-related dropout rates, including decoder dropout, attention dropout, and activation dropout, are set to zero.
Instead, regularization is achieved through metadata and text dropout with a rate of 0.2, together with label smoothing.
Models are optimized using Adam with betas set to 0.9 and 0.98.
We adopt a tri-stage learning rate schedule with warmup, hold, and decay phases, using a base learning rate of $4 \times 10^{-5}$ and 800 warmup steps.
The batch size is controlled by a maximum of 4096 target tokens, and training runs for 40 to 80 epochs depending on the experiment.
At inference time, we use greedy decoding or beam search with a beam size of 5.

\subsection{Baseline Evaluation}
\label{sec:baseline_eval}

\begin{figure}[t]
  \centering
  \includegraphics[width=1.05\columnwidth]{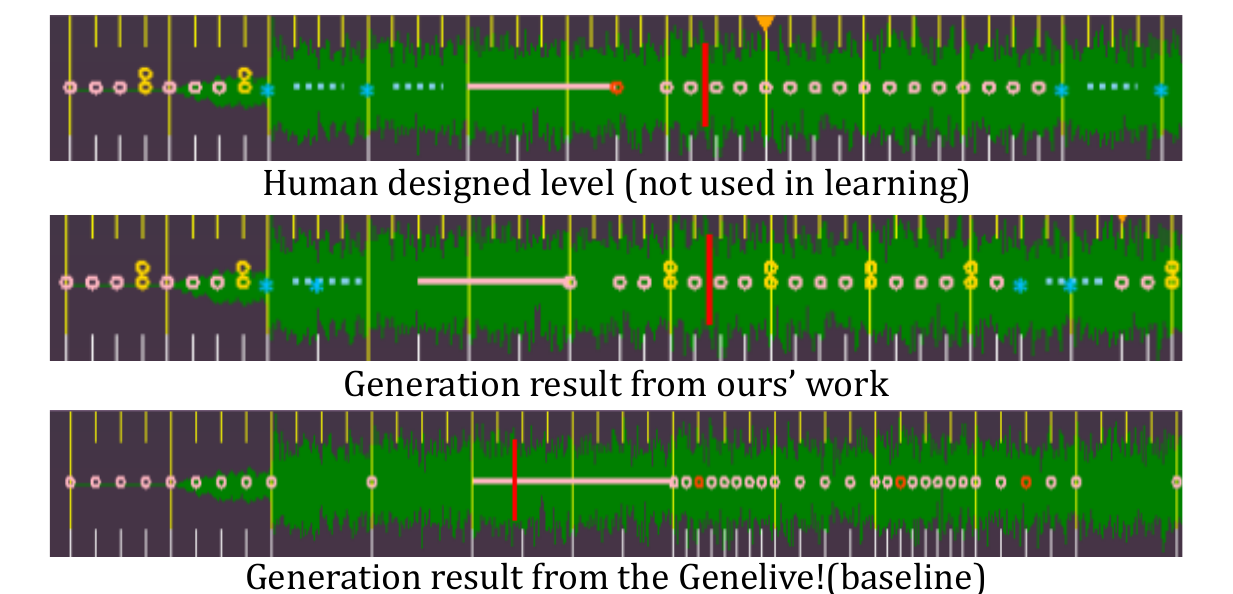}
  \caption{Representative generation examples under identical audio and metadata conditions.
  From top to bottom: a human-designed level segment (not used in training), the output of the proposed model, and the output of a frame-based baseline.
  All examples are aligned in beat space and cover the same musical excerpt, allowing direct comparison of event spacing and timing structure.}
  \Description{Three horizontal gameplay strips stacked vertically, each showing the same musical excerpt along a shared beat timeline. Every strip renders the audio waveform as a green band behind the level and places colored markers on top: circle markers for taps, asterisk markers for breaks, short horizontal bars for holds, and a vertical red bar for a double. Yellow vertical lines indicate beat divisions. The top strip is labeled ``Human designed level (not used in learning)'' and contains densely grouped taps near the start, a short hold mid-segment, and a run of evenly spaced taps toward the end. The middle strip, labeled ``Generation result from ours' work'', exhibits similar event density and spacing, closely tracking the human chart's rhythmic structure. The bottom strip, labeled ``Generation result from the Genelive!(baseline)'', shows sparser taps with less clear rhythmic grouping and a longer, offset hold, illustrating the frame-based baseline's weaker event-level alignment.}
  \label{fig:case_example}
\end{figure}
\begin{table*}[t]
\centering
\caption{
Baseline evaluation across difficulty settings.
We report performance using two metrics.
Event-level F1 is the primary metric, evaluating multi-class gameplay event prediction after reconstruction, with a temporal tolerance of $\frac{1}{8}$ beat.
Frame-level F1 is reported as a diagnostic metric for binary onset detection at the frame level.
For frame-level baselines (DDC, GeneLive!), Frame-F1 is computed on temporally smoothed predictions before peak-based event reconstruction.
For our sequence-to-sequence model, Frame-level F1 (Onset) is computed by rasterizing generated events to frame-wise onsets (optionally with light temporal smoothing) for comparison.
}
\label{tab:baseline_pretty}
\setlength{\tabcolsep}{6pt}
\renewcommand{\arraystretch}{1.2}
\begin{tabular}{l cc cc cc}
\toprule
& \multicolumn{2}{c}{\textbf{DDC}} & \multicolumn{2}{c}{\textbf{GeneLive!}} & \multicolumn{2}{c}{\textbf{Ours}} \\
\cmidrule(lr){2-3} \cmidrule(lr){4-5} \cmidrule(lr){6-7}
\textbf{Difficulty} & \textbf{Event} & \textbf{Onset} & \textbf{Event} & \textbf{Onset} & \textbf{Event} & \textbf{Onset} \\
\midrule
Beginner  & 0.337 & 0.728 & 0.392 & 0.784 & \textbf{0.651} & \textbf{0.819} \\
Advanced  & 0.274 & 0.692 & 0.324 & 0.722 & \textbf{0.575} & \textbf{0.728} \\
Expert    & 0.231 & 0.664 & 0.275 & 0.696 & \textbf{0.526} & \textbf{0.774} \\
Master    & 0.219 & 0.682 & 0.246 & \textbf{0.711} & \textbf{0.481} & 0.692 \\
Remaster  & 0.187 & 0.652 & 0.215 & \textbf{0.687} & \textbf{0.457} & 0.580 \\
\midrule
\textit{Average} & 0.254 & 0.686 & 0.298 & 0.734 & \textbf{0.527} & \textbf{0.733} \\
\bottomrule
\end{tabular}
\end{table*}
We compare our sequence-based approach with two representative external frame-level baselines, DDC and GeneLive!.
In their original settings, these methods are typically formulated as binary onset detection at the audio-frame level.
In this work, we adapt both baselines to our simai-style vocabulary and train them as a 6-way frame-level classifier over \{none, tap, hold, slide, double, break\}.

We evaluate all methods using event-level metrics that assess the accuracy of predicted gameplay events and their temporal placement.
We report event-level Precision, Recall, and F1-score after mapping every model output to a discrete event sequence.
This mapping provides a common evaluation abstraction across different modeling formulations.
For the adapted frame-level baselines, per-frame 6-class predictions are smoothed temporally and then converted to discrete events via local peak detection.
This conversion yields both event timing and event type under the same label set.
For our method, decoded beat-shift and event-type token sequences are deterministically reconstructed into events.
Predicted events are aligned to ground-truth events using a temporal tolerance of $\frac{1}{8}$ beat to account for minor timing deviations.

Table~\ref{tab:baseline_pretty} reports event-level F1-scores across difficulty settings.
For DDC and GeneLive!, we additionally report an onset-only frame-level F1 score.
It evaluates binary onset detection before peak-based event reconstruction.
This onset-level score is included only as a reference for timing accuracy under the original binary-onset formulation.
It is not directly comparable to the event-level F1 used as the primary metric in this work.
We observe that the adapted baselines retain reasonable onset detection performance on this dataset.
Their substantially lower event-level F1 therefore reflects the mismatch between a frame-level onset formulation and event-level reconstruction requirements under a multi-class action vocabulary.
As illustrated in Figure~\ref{fig:case_example}, this mismatch is reflected in the generated event spacing and timing patterns.
All baseline models are retrained on the same training split of the maimai dataset using publicly released implementations.
Unless otherwise stated, we use the default training settings provided by the original implementations.
Our comparison focuses on contrasting modeling paradigms under the same event-level evaluation protocol.

\subsection{Model Variants}
\label{sec:model_variants}

We further analyze several internal variants of our sequence-based model to examine how different design choices affect generation quality.
To better isolate the role of audio in this analysis, we introduce the Audio Contribution Score (ACS).
Even without strong local audio cues, a model can exploit statistical regularities in the token sequence that are already captured by metadata conditioning and by the autoregressive prior learned during training, so raw performance differences between audio-enabled settings are difficult to attribute to audio alone.

ACS quantifies how much of the model's performance depends on audio beyond these non-audio factors by normalizing the performance gain obtained from correct audio input.
Specifically, for a given evaluation score $S$, we define
\[
\mathrm{ACS}
=
\frac{S_{\mathrm{true}} - S_{\mathrm{shuf}}}
     {S_{\mathrm{forced}} - S_{\mathrm{shuf}}},
\]
where $S_{\mathrm{true}}$ denotes the score obtained using the correct audio input under standard autoregressive decoding,
$S_{\mathrm{shuf}}$ denotes the score obtained when audio inputs are randomly shuffled across samples at evaluation time,
and $S_{\mathrm{forced}}$ denotes a teacher-forced reference score obtained by decoding with ground-truth previous tokens while retaining the same audio and metadata inputs.

Importantly, $S_{\mathrm{forced}}$ does not represent a theoretical maximum or an assumption of a unique correct chart.
Instead, it reflects an empirical upper bound under the current model and evaluation protocol, in which exposure errors from autoregressive decoding are removed.
This normalization isolates the relative contribution of meaningful audio--event alignment beyond metadata conditioning and sequence priors.
An ACS value close to zero indicates that performance is largely explained by non-audio factors, while higher ACS values indicate stronger dependence on audio for accurate event placement.

Table~\ref{tab:main_results} summarizes the results.
When metadata is removed, ACS increases, reflecting the fact that audio contributes a larger relative share of the remaining predictive signal under reduced non-audio conditioning.
This behavior is expected, as fewer non-audio cues are available to explain event timing and density.
We also compare different audio encoders.
Models using the Whisper encoder consistently outperform those using MERT under the same event-level evaluation protocol.
This difference can be attributed, in part, to the higher temporal resolution of Whisper’s acoustic representations
(approximately 20--25\,ms per frame, compared to 40--50\,ms for MERT),
which provides finer-grained timing cues for event placement within the fixed temporal tolerance used in evaluation.

\begin{table}[t]
\centering
\caption{Main results for sequence-based models and internal variants.}
\label{tab:main_results}
\begin{tabular}{l c c}
\toprule
Model & Event F1$\uparrow$ & ACS$\uparrow$ \\
\midrule
Whisper (full)       & 0.527 & 0.58 \\
Whisper (zero-meta) & 0.498 & 0.67 \\
zero-audio & 0.266 &  - \\
MERT encoder        & 0.502 & 0.54 \\
\bottomrule
\end{tabular}
\end{table}

\subsection{Observed Failure Modes under Challenging Conditions}
\label{sec:failure_modes}

While the proposed model demonstrates stable performance under typical chart settings, we observe several recurring failure modes under more challenging conditions.

\paragraph{Quantifying degeneracy.}
To quantify two common degeneracies discussed below, we report
(i) \emph{extreme-density rate}, the fraction of fixed-length beat windows whose event count exceeds a density threshold,
and (ii) \emph{loop-collapse rate}, the fraction of segments that contain an exactly repeated token pattern of length at least $L$ repeated consecutively at least $R$ times.
Unless otherwise stated, we use 8-beat windows and count events per window.
For the density threshold, we set $K$ to the 99th percentile of window event counts on the validation set (i.e., \emph{top 1\%} densest windows under a reference distribution), which yields $K=40$ events per 8 beats in our data.
For loop-collapse, we set $L=16$ tokens and $R=4$ repeats, and apply the test to the model output token sequence.
Table~\ref{tab:degeneracy} summarizes these rates (lower is better).

\begin{table}[!htbp]
\centering
\caption{Degeneracy diagnostics (lower is better). Extreme-density is the percentage of 8-beat windows exceeding $K=40$ events (set as the 99th percentile on validation); loop-collapse is the percentage of segments containing a repeated pattern of length $\ge L=16$ tokens repeated $\ge R=4$ times.}
\label{tab:degeneracy}
\begin{tabular}{lcc}
\toprule
Model & Extreme-density (\%)$\downarrow$ & Loop-collapse (\%)$\downarrow$ \\
\midrule
Whisper (full)      & 2.1 & 0.4 \\
Whisper (zero-meta) & 3.8 & 1.7 \\
zero-audio          & 5.9 & 4.3 \\
\bottomrule
\end{tabular}
\end{table}

First, although global rhythmic alignment is generally preserved, onset tokens occasionally deviate from exact beat positions in transition segments. This behavior is particularly noticeable when long beat-shift intervals are involved, suggesting an accumulation effect in autoregressive decoding that manifests as phase shift over extended temporal gaps.

Second, at higher difficulty levels, generated levels tend to exhibit limited diversity in event types. While overall note density increases as expected, complex interaction patterns characteristic of high-level levels remain underrepresented. This indicates a ceiling imposed by the current event vocabulary, which captures core interaction primitives but does not fully represent more complex composite interactions.

Third, under high-BPM conditions, we observe cases where generated event sequences violate ergonomic constraints of the physical arcade interface, resulting in physically conflicting actions. This highlights a limitation of purely symbolic sequence modeling, which does not explicitly account for spatial or physical constraints inherent to the gameplay device.

Finally, in some segments, generated events appear weakly anchored to specific instrumental cues, making it difficult to identify which musical component is being followed. This observation is consistent with our encoder ablation results, suggesting that the resolution and structure of audio representations directly influence the model's ability to track instrument-specific transients.

Together, these observations delineate the current modeling boundaries of token-level autoregressive generation and point to promising directions for future work, including richer event vocabularies, spatially aware representations, and more instrument-sensitive audio conditioning.

\section{Conclusion}

This paper introduced an event-based sequence-to-sequence formulation for music game level modeling, representing gameplay levels as discrete token sequences composed of beat-shift and gameplay-event tokens in beat space.
By conditioning autoregressive event prediction on both audio features and level metadata, the proposed multimodal model makes music-to-event correspondence explicit at the event level, beyond frame-based temporal grids.
Experimental results demonstrate consistent improvements over frame-level baselines under event-level evaluation metrics, while analysis suggests a division of roles between input modalities: audio provides cues for local temporal alignment, whereas metadata such as BPM and difficulty primarily governs higher-level structural tendencies.
To support systematic analysis of this interaction, we proposed the Audio Contribution Score (ACS) as a diagnostic metric that quantifies the marginal effect of audio conditioning under a unified evaluation protocol.
Future work may extend this event-based framework with positional attributes, longer-context modeling, and evaluation protocols that better reflect perceptual and playability constraints.

\bibliographystyle{ACM-Reference-Format}
\bibliography{refs}


\begin{thebibliography}{32}


\ifx \showCODEN    \undefined \def \showCODEN     #1{\unskip}     \fi
\ifx \showDOI      \undefined \def \showDOI       #1{#1}\fi
\ifx \showISBNx    \undefined \def \showISBNx     #1{\unskip}     \fi
\ifx \showISBNxiii \undefined \def \showISBNxiii  #1{\unskip}     \fi
\ifx \showISSN     \undefined \def \showISSN      #1{\unskip}     \fi
\ifx \showLCCN     \undefined \def \showLCCN      #1{\unskip}     \fi
\ifx \shownote     \undefined \def \shownote      #1{#1}          \fi
\ifx \showarticletitle \undefined \def \showarticletitle #1{#1}   \fi
\ifx \showURL      \undefined \def \showURL       {\relax}        \fi
\providecommand\bibfield[2]{#2}
\providecommand\bibinfo[2]{#2}
\providecommand\natexlab[1]{#1}
\providecommand\showeprint[2][]{arXiv:#2}

\bibitem[Bengio et~al\mbox{.}(2013)]%
        {bengio2013representation}
\bibfield{author}{\bibinfo{person}{Yoshua Bengio}, \bibinfo{person}{Aaron
  Courville}, {and} \bibinfo{person}{Pascal Vincent}.}
  \bibinfo{year}{2013}\natexlab{}.
\newblock \showarticletitle{Representation Learning: A Review and New
  Perspectives}.
\newblock \bibinfo{journal}{\emph{IEEE Transactions on Pattern Analysis and
  Machine Intelligence}} \bibinfo{volume}{35}, \bibinfo{number}{8}
  (\bibinfo{year}{2013}), \bibinfo{pages}{1798--1828}.
\newblock
\urldef\tempurl%
\url{https://doi.org/10.1109/TPAMI.2013.50}
\showDOI{\tempurl}


\bibitem[Borsos et~al\mbox{.}(2023)]%
        {borsos2022audiolm}
\bibfield{author}{\bibinfo{person}{Zal{\'a}n Borsos},
  \bibinfo{person}{Rapha{\"e}l Marinier}, \bibinfo{person}{Damien Vincent},
  \bibinfo{person}{Eugene Kharitonov}, \bibinfo{person}{Olivier Pietquin},
  \bibinfo{person}{Matthew Sharifi}, \bibinfo{person}{Dominik Roblek},
  \bibinfo{person}{Olivier Teboul}, \bibinfo{person}{David Grangier},
  \bibinfo{person}{Marco Tagliasacchi}, {and} \bibinfo{person}{Neil
  Zeghidour}.} \bibinfo{year}{2023}\natexlab{}.
\newblock \showarticletitle{{AudioLM}: A Language Modeling Approach to Audio
  Generation}.
\newblock \bibinfo{journal}{\emph{IEEE/ACM Transactions on Audio, Speech, and
  Language Processing}}  \bibinfo{volume}{31} (\bibinfo{year}{2023}),
  \bibinfo{pages}{2523--2533}.
\newblock
\urldef\tempurl%
\url{https://doi.org/10.1109/TASLP.2023.3288409}
\showDOI{\tempurl}


\bibitem[Boulanger-Lewandowski et~al\mbox{.}(2012)]%
        {boulanger2012modeling}
\bibfield{author}{\bibinfo{person}{Nicolas Boulanger-Lewandowski},
  \bibinfo{person}{Yoshua Bengio}, {and} \bibinfo{person}{Pascal Vincent}.}
  \bibinfo{year}{2012}\natexlab{}.
\newblock \bibinfo{title}{Modeling Temporal Dependencies in High-Dimensional
  Sequences: Application to Polyphonic Music Generation and Transcription}.
\newblock
\newblock
\showeprint[arxiv]{1206.6392}~[cs.LG]
\urldef\tempurl%
\url{https://arxiv.org/abs/1206.6392}
\showURL{%
\tempurl}


\bibitem[Casey et~al\mbox{.}(2008)]%
        {casey2008content}
\bibfield{author}{\bibinfo{person}{Michael~A. Casey}, \bibinfo{person}{Remco~C.
  Veltkamp}, \bibinfo{person}{Masataka Goto}, \bibinfo{person}{Marc Leman},
  \bibinfo{person}{Christophe Rhodes}, {and} \bibinfo{person}{Malcolm Slaney}.}
  \bibinfo{year}{2008}\natexlab{}.
\newblock \showarticletitle{Content-based Music Information Retrieval: Current
  Directions and Future Challenges}.
\newblock \bibinfo{journal}{\emph{Proc. IEEE}} \bibinfo{volume}{96},
  \bibinfo{number}{4} (\bibinfo{year}{2008}), \bibinfo{pages}{668--696}.
\newblock
\urldef\tempurl%
\url{https://doi.org/10.1109/JPROC.2008.916370}
\showDOI{\tempurl}


\bibitem[Chan et~al\mbox{.}(2015)]%
        {chan2015las}
\bibfield{author}{\bibinfo{person}{William Chan}, \bibinfo{person}{Navdeep
  Jaitly}, \bibinfo{person}{Quoc~V. Le}, {and} \bibinfo{person}{Oriol
  Vinyals}.} \bibinfo{year}{2015}\natexlab{}.
\newblock \bibinfo{title}{Listen, Attend and Spell}.
\newblock
\newblock
\showeprint[arxiv]{1508.01211}~[cs.CL]
\urldef\tempurl%
\url{https://arxiv.org/abs/1508.01211}
\showURL{%
\tempurl}


\bibitem[Chan et~al\mbox{.}(2016)]%
        {chan2016listen}
\bibfield{author}{\bibinfo{person}{William Chan}, \bibinfo{person}{Navdeep
  Jaitly}, \bibinfo{person}{Quoc~V. Le}, {and} \bibinfo{person}{Oriol
  Vinyals}.} \bibinfo{year}{2016}\natexlab{}.
\newblock \showarticletitle{Listen, Attend and Spell: A Neural Network for
  Large Vocabulary Conversational Speech Recognition}. In
  \bibinfo{booktitle}{\emph{2016 IEEE International Conference on Acoustics,
  Speech and Signal Processing (ICASSP)}}. \bibinfo{publisher}{IEEE},
  \bibinfo{pages}{4960--4964}.
\newblock
\urldef\tempurl%
\url{https://doi.org/10.1109/ICASSP.2016.7472621}
\showDOI{\tempurl}


\bibitem[Devlin et~al\mbox{.}(2019)]%
        {devlin2019bert}
\bibfield{author}{\bibinfo{person}{Jacob Devlin}, \bibinfo{person}{Ming-Wei
  Chang}, \bibinfo{person}{Kenton Lee}, {and} \bibinfo{person}{Kristina
  Toutanova}.} \bibinfo{year}{2019}\natexlab{}.
\newblock \showarticletitle{{BERT}: Pre-training of Deep Bidirectional
  Transformers for Language Understanding}. In
  \bibinfo{booktitle}{\emph{Proceedings of the 2019 Conference of the North
  American Chapter of the Association for Computational Linguistics: Human
  Language Technologies (NAACL-HLT)}}. \bibinfo{publisher}{Association for
  Computational Linguistics}, \bibinfo{pages}{4171--4186}.
\newblock
\urldef\tempurl%
\url{https://doi.org/10.18653/v1/N19-1423}
\showDOI{\tempurl}


\bibitem[Donahue et~al\mbox{.}(2017)]%
        {donahue2017dance}
\bibfield{author}{\bibinfo{person}{Chris Donahue}, \bibinfo{person}{Zachary~C.
  Lipton}, {and} \bibinfo{person}{Julian McAuley}.}
  \bibinfo{year}{2017}\natexlab{}.
\newblock \showarticletitle{Dance Dance Convolution}. In
  \bibinfo{booktitle}{\emph{Proceedings of the 34th International Conference on
  Machine Learning (ICML)}} \emph{(\bibinfo{series}{Proceedings of Machine
  Learning Research}, Vol.~\bibinfo{volume}{70})}. \bibinfo{publisher}{PMLR},
  \bibinfo{pages}{1934--1943}.
\newblock
\urldef\tempurl%
\url{http://proceedings.mlr.press/v70/donahue17a.html}
\showURL{%
\tempurl}


\bibitem[Dong et~al\mbox{.}(2018)]%
        {dong2018musegan}
\bibfield{author}{\bibinfo{person}{Hao-Wen Dong}, \bibinfo{person}{Wen-Yi
  Hsiao}, \bibinfo{person}{Li-Chia Yang}, {and} \bibinfo{person}{Yi-Hsuan
  Yang}.} \bibinfo{year}{2018}\natexlab{}.
\newblock \showarticletitle{{MuseGAN}: Multi-track Sequential Generative
  Adversarial Networks for Symbolic Music Generation and Accompaniment}. In
  \bibinfo{booktitle}{\emph{Proceedings of the AAAI Conference on Artificial
  Intelligence}}, Vol.~\bibinfo{volume}{32}. \bibinfo{publisher}{AAAI Press},
  \bibinfo{pages}{34--41}.
\newblock
\urldef\tempurl%
\url{https://doi.org/10.1609/aaai.v32i1.11312}
\showDOI{\tempurl}


\bibitem[Ellis(2007)]%
        {ellis2007beat}
\bibfield{author}{\bibinfo{person}{Daniel P.~W. Ellis}.}
  \bibinfo{year}{2007}\natexlab{}.
\newblock \showarticletitle{Beat Tracking by Dynamic Programming}.
\newblock \bibinfo{journal}{\emph{Journal of New Music Research}}
  \bibinfo{volume}{36}, \bibinfo{number}{1} (\bibinfo{year}{2007}),
  \bibinfo{pages}{51--60}.
\newblock
\urldef\tempurl%
\url{https://doi.org/10.1080/09298210701653344}
\showDOI{\tempurl}


\bibitem[Goto(2017)]%
        {goto2017miroverview}
\bibfield{author}{\bibinfo{person}{Masataka Goto}.}
  \bibinfo{year}{2017}\natexlab{}.
\newblock \bibinfo{title}{A Review of Music Information Retrieval Techniques}.
\newblock
\newblock
\showeprint[arxiv]{1709.04396}~[cs.SD]
\urldef\tempurl%
\url{https://arxiv.org/abs/1709.04396}
\showURL{%
\tempurl}


\bibitem[Goto and Muraoka(2001)]%
        {goto2001realtime}
\bibfield{author}{\bibinfo{person}{Masataka Goto} {and} \bibinfo{person}{Yoichi
  Muraoka}.} \bibinfo{year}{2001}\natexlab{}.
\newblock \showarticletitle{An Audio-based Real-time Beat Tracking System for
  Music with or without Drum-sounds}.
\newblock \bibinfo{journal}{\emph{Journal of New Music Research}}
  \bibinfo{volume}{30}, \bibinfo{number}{2} (\bibinfo{year}{2001}),
  \bibinfo{pages}{159--171}.
\newblock
\urldef\tempurl%
\url{https://doi.org/10.1076/jnmr.30.2.159.7119}
\showDOI{\tempurl}


\bibitem[Graves et~al\mbox{.}(2006)]%
        {graves2006ctc}
\bibfield{author}{\bibinfo{person}{Alex Graves}, \bibinfo{person}{Santiago
  Fern{\'a}ndez}, \bibinfo{person}{Faustino Gomez}, {and}
  \bibinfo{person}{J{\"u}rgen Schmidhuber}.} \bibinfo{year}{2006}\natexlab{}.
\newblock \showarticletitle{Connectionist Temporal Classification: Labelling
  Unsegmented Sequence Data with Recurrent Neural Networks}. In
  \bibinfo{booktitle}{\emph{Proceedings of the 23rd International Conference on
  Machine Learning (ICML)}}. \bibinfo{publisher}{ACM},
  \bibinfo{pages}{369--376}.
\newblock
\urldef\tempurl%
\url{https://doi.org/10.1145/1143844.1143891}
\showDOI{\tempurl}


\bibitem[Hadjeres et~al\mbox{.}(2017)]%
        {hadjeres2017deepbach}
\bibfield{author}{\bibinfo{person}{Ga{\"e}tan Hadjeres},
  \bibinfo{person}{Fran{\c{c}}ois Pachet}, {and} \bibinfo{person}{Frank
  Nielsen}.} \bibinfo{year}{2017}\natexlab{}.
\newblock \showarticletitle{{DeepBach}: A Steerable Model for {Bach} Chorales
  Generation}. In \bibinfo{booktitle}{\emph{Proceedings of the 34th
  International Conference on Machine Learning (ICML)}}
  \emph{(\bibinfo{series}{Proceedings of Machine Learning Research},
  Vol.~\bibinfo{volume}{70})}. \bibinfo{publisher}{PMLR},
  \bibinfo{pages}{1362--1371}.
\newblock
\urldef\tempurl%
\url{http://proceedings.mlr.press/v70/hadjeres17a.html}
\showURL{%
\tempurl}


\bibitem[Hochreiter and Schmidhuber(1997)]%
        {hochreiter1997lstm}
\bibfield{author}{\bibinfo{person}{Sepp Hochreiter} {and}
  \bibinfo{person}{J{\"u}rgen Schmidhuber}.} \bibinfo{year}{1997}\natexlab{}.
\newblock \showarticletitle{Long Short-Term Memory}.
\newblock \bibinfo{journal}{\emph{Neural Computation}} \bibinfo{volume}{9},
  \bibinfo{number}{8} (\bibinfo{year}{1997}), \bibinfo{pages}{1735--1780}.
\newblock
\urldef\tempurl%
\url{https://doi.org/10.1162/neco.1997.9.8.1735}
\showDOI{\tempurl}


\bibitem[Huang et~al\mbox{.}(2017)]%
        {huang2016counterpoint}
\bibfield{author}{\bibinfo{person}{Cheng-Zhi~Anna Huang}, \bibinfo{person}{Tim
  Cooijmans}, \bibinfo{person}{Adam Roberts}, \bibinfo{person}{Aaron
  Courville}, {and} \bibinfo{person}{Douglas Eck}.}
  \bibinfo{year}{2017}\natexlab{}.
\newblock \showarticletitle{Counterpoint by Convolution}. In
  \bibinfo{booktitle}{\emph{Proceedings of the 18th International Society for
  Music Information Retrieval Conference (ISMIR)}}. \bibinfo{pages}{211--218}.
\newblock
\urldef\tempurl%
\url{https://ismir2017.smcnus.org/wp-content/uploads/2017/10/187_Paper.pdf}
\showURL{%
\tempurl}


\bibitem[Huang et~al\mbox{.}(2019)]%
        {huang2018music}
\bibfield{author}{\bibinfo{person}{Cheng-Zhi~Anna Huang},
  \bibinfo{person}{Ashish Vaswani}, \bibinfo{person}{Jakob Uszkoreit},
  \bibinfo{person}{Noam Shazeer}, \bibinfo{person}{Ian Simon},
  \bibinfo{person}{Curtis Hawthorne}, \bibinfo{person}{Andrew~M. Dai},
  \bibinfo{person}{Matthew~D. Hoffman}, \bibinfo{person}{Monica Dinculescu},
  {and} \bibinfo{person}{Douglas Eck}.} \bibinfo{year}{2019}\natexlab{}.
\newblock \showarticletitle{Music Transformer: Generating Music with Long-Term
  Structure}. In \bibinfo{booktitle}{\emph{International Conference on Learning
  Representations (ICLR)}}.
\newblock
\urldef\tempurl%
\url{https://openreview.net/forum?id=rJe4ShAcF7}
\showURL{%
\tempurl}


\bibitem[Huang and Yang(2020)]%
        {huang2020pop}
\bibfield{author}{\bibinfo{person}{Yu-Siang Huang} {and}
  \bibinfo{person}{Yi-Hsuan Yang}.} \bibinfo{year}{2020}\natexlab{}.
\newblock \showarticletitle{Pop Music Transformer: Beat-based Modeling and
  Generation of Expressive Pop Piano Compositions}. In
  \bibinfo{booktitle}{\emph{Proceedings of the 28th ACM International
  Conference on Multimedia (MM)}}. \bibinfo{publisher}{ACM},
  \bibinfo{pages}{1180--1188}.
\newblock
\urldef\tempurl%
\url{https://doi.org/10.1145/3394171.3413671}
\showDOI{\tempurl}


\bibitem[Li et~al\mbox{.}(2024)]%
        {li2024mert}
\bibfield{author}{\bibinfo{person}{Yizhi Li}, \bibinfo{person}{Ruibin Yuan},
  \bibinfo{person}{Ge Zhang}, \bibinfo{person}{Yinghao Ma},
  \bibinfo{person}{Xingran Chen}, \bibinfo{person}{Hanzhi Yin},
  \bibinfo{person}{Chenghua Xiao}, \bibinfo{person}{Chenghua Lin},
  \bibinfo{person}{Anton Ragni}, \bibinfo{person}{Emmanouil Benetos},
  \bibinfo{person}{Norbert Gyenge}, \bibinfo{person}{Roger Dannenberg},
  \bibinfo{person}{Ruibo Liu}, \bibinfo{person}{Wenhu Chen},
  \bibinfo{person}{Gus Xia}, \bibinfo{person}{Yemin Shi},
  \bibinfo{person}{Wenhao Huang}, \bibinfo{person}{Zili Wang},
  \bibinfo{person}{Yike Guo}, {and} \bibinfo{person}{Jie Fu}.}
  \bibinfo{year}{2024}\natexlab{}.
\newblock \bibinfo{title}{{MERT}: Acoustic Music Understanding Model with
  Large-Scale Self-supervised Training}.
\newblock
\newblock
\showeprint[arxiv]{2306.00107}~[cs.SD]
\urldef\tempurl%
\url{https://arxiv.org/abs/2306.00107}
\showURL{%
\tempurl}


\bibitem[Liang et~al\mbox{.}(2019)]%
        {liang2019procedural}
\bibfield{author}{\bibinfo{person}{Yubin Liang}, \bibinfo{person}{Wanxiang Li},
  {and} \bibinfo{person}{Kokolo Ikeda}.} \bibinfo{year}{2019}\natexlab{}.
\newblock \showarticletitle{Procedural Content Generation of Rhythm Games Using
  Deep Learning Methods}. In \bibinfo{booktitle}{\emph{Entertainment Computing
  and Serious Games (ICEC-JCSG 2019)}} \emph{(\bibinfo{series}{Lecture Notes in
  Computer Science}, Vol.~\bibinfo{volume}{11863})}.
  \bibinfo{publisher}{Springer}, \bibinfo{address}{Cham},
  \bibinfo{pages}{134--145}.
\newblock
\urldef\tempurl%
\url{https://doi.org/10.1007/978-3-030-34644-7_11}
\showDOI{\tempurl}


\bibitem[Lin et~al\mbox{.}(2019)]%
        {lin2019generationmania}
\bibfield{author}{\bibinfo{person}{Zhiyu Lin}, \bibinfo{person}{Kyle Xiao},
  {and} \bibinfo{person}{Mark Riedl}.} \bibinfo{year}{2019}\natexlab{}.
\newblock \showarticletitle{{GenerationMania}: Learning to Semantically
  Choreograph}. In \bibinfo{booktitle}{\emph{Proceedings of the AAAI Conference
  on Artificial Intelligence and Interactive Digital Entertainment (AIIDE)}},
  Vol.~\bibinfo{volume}{15}. \bibinfo{publisher}{AAAI Press},
  \bibinfo{pages}{52--58}.
\newblock
\urldef\tempurl%
\url{https://ojs.aaai.org/index.php/AIIDE/article/view/5224}
\showURL{%
\tempurl}


\bibitem[Oore et~al\mbox{.}(2020)]%
        {oore2018time}
\bibfield{author}{\bibinfo{person}{Sageev Oore}, \bibinfo{person}{Ian Simon},
  \bibinfo{person}{Sander Dieleman}, \bibinfo{person}{Douglas Eck}, {and}
  \bibinfo{person}{Karen Simonyan}.} \bibinfo{year}{2020}\natexlab{}.
\newblock \showarticletitle{This Time with Feeling: Learning Expressive Musical
  Performance}.
\newblock \bibinfo{journal}{\emph{Neural Computing and Applications}}
  \bibinfo{volume}{32}, \bibinfo{number}{4} (\bibinfo{year}{2020}),
  \bibinfo{pages}{955--967}.
\newblock
\urldef\tempurl%
\url{https://doi.org/10.1007/s00521-018-3758-9}
\showDOI{\tempurl}


\bibitem[Payne(2019)]%
        {payne2019musenet}
\bibfield{author}{\bibinfo{person}{Christine Payne}.}
  \bibinfo{year}{2019}\natexlab{}.
\newblock \bibinfo{title}{{MuseNet}}.
\newblock \bibinfo{howpublished}{OpenAI Blog}.
\newblock
\urldef\tempurl%
\url{https://openai.com/blog/musenet}
\showURL{%
\tempurl}
\newblock
\shownote{Accessed: 2026-01-29}.


\bibitem[Radford et~al\mbox{.}(2023)]%
        {radford2022whisper}
\bibfield{author}{\bibinfo{person}{Alec Radford}, \bibinfo{person}{Jong~Wook
  Kim}, \bibinfo{person}{Tao Xu}, \bibinfo{person}{Greg Brockman},
  \bibinfo{person}{Christine McLeavey}, {and} \bibinfo{person}{Ilya
  Sutskever}.} \bibinfo{year}{2023}\natexlab{}.
\newblock \showarticletitle{Robust Speech Recognition via Large-Scale Weak
  Supervision}. In \bibinfo{booktitle}{\emph{Proceedings of the 40th
  International Conference on Machine Learning (ICML)}}
  \emph{(\bibinfo{series}{Proceedings of Machine Learning Research},
  Vol.~\bibinfo{volume}{202})}. \bibinfo{publisher}{PMLR},
  \bibinfo{pages}{28492--28518}.
\newblock
\urldef\tempurl%
\url{https://proceedings.mlr.press/v202/radford23a.html}
\showURL{%
\tempurl}


\bibitem[Roberts et~al\mbox{.}(2018)]%
        {roberts2018hierarchical}
\bibfield{author}{\bibinfo{person}{Adam Roberts}, \bibinfo{person}{Jesse
  Engel}, \bibinfo{person}{Colin Raffel}, \bibinfo{person}{Curtis Hawthorne},
  {and} \bibinfo{person}{Douglas Eck}.} \bibinfo{year}{2018}\natexlab{}.
\newblock \showarticletitle{A Hierarchical Latent Vector Model for Learning
  Long-Term Structure in Music}. In \bibinfo{booktitle}{\emph{Proceedings of
  the 35th International Conference on Machine Learning (ICML)}}
  \emph{(\bibinfo{series}{Proceedings of Machine Learning Research},
  Vol.~\bibinfo{volume}{80})}. \bibinfo{publisher}{PMLR},
  \bibinfo{pages}{4364--4373}.
\newblock
\urldef\tempurl%
\url{http://proceedings.mlr.press/v80/roberts18a.html}
\showURL{%
\tempurl}


\bibitem[Schedl et~al\mbox{.}(2014)]%
        {schedl2014mir}
\bibfield{author}{\bibinfo{person}{Markus Schedl}, \bibinfo{person}{Emilia
  G{\'o}mez}, {and} \bibinfo{person}{Juli{\'a}n Urbano}.}
  \bibinfo{year}{2014}\natexlab{}.
\newblock \showarticletitle{Music Information Retrieval: Recent Developments
  and Applications}.
\newblock \bibinfo{journal}{\emph{Foundations and Trends in Information
  Retrieval}} \bibinfo{volume}{8}, \bibinfo{number}{2-3}
  (\bibinfo{year}{2014}), \bibinfo{pages}{127--261}.
\newblock
\urldef\tempurl%
\url{https://doi.org/10.1561/1500000042}
\showDOI{\tempurl}


\bibitem[{simai Community}(2020)]%
        {simai_maidata_spec}
\bibfield{author}{\bibinfo{person}{{simai Community}}.}
  \bibinfo{year}{2020}\natexlab{}.
\newblock \bibinfo{title}{Maidata Specification for {StepMania}-Style Charts}.
\newblock
  \bibinfo{howpublished}{\url{https://w.atwiki.jp/simai/pages/1003.html}}.
\newblock
\newblock
\shownote{Community specification; accessed 2026-01-29}.


\bibitem[Srivastava and Salakhutdinov(2012)]%
        {srivastava2012multimodal}
\bibfield{author}{\bibinfo{person}{Nitish Srivastava} {and}
  \bibinfo{person}{Ruslan Salakhutdinov}.} \bibinfo{year}{2012}\natexlab{}.
\newblock \showarticletitle{Multimodal Learning with Deep {Boltzmann}
  Machines}. In \bibinfo{booktitle}{\emph{Advances in Neural Information
  Processing Systems (NeurIPS)}}, Vol.~\bibinfo{volume}{25}.
  \bibinfo{publisher}{Curran Associates}, \bibinfo{pages}{2222--2230}.
\newblock
\urldef\tempurl%
\url{https://proceedings.neurips.cc/paper/2012/hash/af21d0c97db2e27e13572cbf59eb343d-Abstract.html}
\showURL{%
\tempurl}


\bibitem[Summerville et~al\mbox{.}(2018)]%
        {summerville2018pcgml}
\bibfield{author}{\bibinfo{person}{Adam Summerville}, \bibinfo{person}{Sam
  Snodgrass}, \bibinfo{person}{Matthew Guzdial}, \bibinfo{person}{Christoffer
  Holmg{\aa}rd}, \bibinfo{person}{Amy~K. Hoover}, \bibinfo{person}{Aaron
  Isaksen}, \bibinfo{person}{Andy Nealen}, {and} \bibinfo{person}{Julian
  Togelius}.} \bibinfo{year}{2018}\natexlab{}.
\newblock \showarticletitle{Procedural Content Generation via Machine Learning
  ({PCGML})}.
\newblock \bibinfo{journal}{\emph{IEEE Transactions on Games}}
  \bibinfo{volume}{10}, \bibinfo{number}{3} (\bibinfo{year}{2018}),
  \bibinfo{pages}{257--270}.
\newblock
\urldef\tempurl%
\url{https://doi.org/10.1109/TG.2018.2846639}
\showDOI{\tempurl}


\bibitem[Takada et~al\mbox{.}(2023)]%
        {takada2023genelive}
\bibfield{author}{\bibinfo{person}{Atsuki Takada}, \bibinfo{person}{Daisuke
  Yamazaki}, \bibinfo{person}{Yusuke Yoshida}, {and}
  \bibinfo{person}{Naranbaatar Ganbat}.} \bibinfo{year}{2023}\natexlab{}.
\newblock \showarticletitle{Gen{\'e}Live!: Generating Rhythm Actions in Love
  Live!}. In \bibinfo{booktitle}{\emph{Proceedings of the AAAI Conference on
  Artificial Intelligence}}, Vol.~\bibinfo{volume}{37}.
  \bibinfo{publisher}{AAAI Press}, \bibinfo{pages}{14477--14484}.
\newblock
\urldef\tempurl%
\url{https://doi.org/10.1609/aaai.v37i11.25657}
\showDOI{\tempurl}


\bibitem[Tsujino and Yamanishi(2018)]%
        {tsujino2018gradation}
\bibfield{author}{\bibinfo{person}{Naoki Tsujino} {and} \bibinfo{person}{Ryohei
  Yamanishi}.} \bibinfo{year}{2018}\natexlab{}.
\newblock \showarticletitle{Dance Dance Gradation: A Generation of Fine-Tuned
  Dance Charts}. In \bibinfo{booktitle}{\emph{Entertainment Computing (ICEC
  2018)}} \emph{(\bibinfo{series}{Lecture Notes in Computer Science},
  Vol.~\bibinfo{volume}{11112})}. \bibinfo{publisher}{Springer},
  \bibinfo{address}{Cham}, \bibinfo{pages}{175--187}.
\newblock
\urldef\tempurl%
\url{https://doi.org/10.1007/978-3-319-99426-0_15}
\showDOI{\tempurl}


\bibitem[Vaswani et~al\mbox{.}(2017)]%
        {vaswani2017attention}
\bibfield{author}{\bibinfo{person}{Ashish Vaswani}, \bibinfo{person}{Noam
  Shazeer}, \bibinfo{person}{Niki Parmar}, \bibinfo{person}{Jakob Uszkoreit},
  \bibinfo{person}{Llion Jones}, \bibinfo{person}{Aidan~N. Gomez},
  \bibinfo{person}{{\L}ukasz Kaiser}, {and} \bibinfo{person}{Illia
  Polosukhin}.} \bibinfo{year}{2017}\natexlab{}.
\newblock \showarticletitle{Attention is All You Need}. In
  \bibinfo{booktitle}{\emph{Advances in Neural Information Processing Systems
  (NeurIPS)}}, Vol.~\bibinfo{volume}{30}. \bibinfo{publisher}{Curran
  Associates}, \bibinfo{pages}{5998--6008}.
\newblock
\urldef\tempurl%
\url{https://arxiv.org/abs/1706.03762}
\showURL{%
\tempurl}


\end{thebibliography}

\end{document}